\documentstyle[11pt,newpasp,twoside,epsfig,graphicx]{article}
\def\edcomment#1{\iffalse\marginpar{\raggedright\sl#1\/}\else\relax\fi}
\reversemarginpar

\begin{document}

\title{The FIRST-APM QSOs Survey (FAQS) in the SBS Region. Preliminary Results}

\author{Vahram Chavushyan, Ra\'ul M\'ujica, Luis Carrasco, Jos\'e R. Vald\'es}

\affil{INAOE. A.P.51 y 216. C.P. 72000. Puebla, Pue., M\'exico  }

\author{Oleg Verkhodanov, and Jivan Stepanian}

\affil{SAO RAS. Karachai-Cherkesia, 357147. Russia.}

\begin{abstract}
The main goal of the FIRST-APM QSO Survey (FAQS) survey is to compile the 
most complete sample of Bright QSOs, located in the area covered by the 
Second Byurakan Survey (SBS). Here we report the first results of an 
ongoing study based on the cross-identification of the FIRST radio catalog 
and the APM optical catalog. The overlapping sky area between  FIRST and SBS  
is about 700 deg$^{2}$. The compiled list of sources for this overlapping 
region contains $\sim 400$ quasar candidates brighter than $B=18\fm5$. About 
90 objects were already spectroscopically classified.  During 1999-2000, we 
observed spectroscopically more than 150 FAQS objects with the 2.1m 
telescope of the Guillermo Haro Observatory (GHO).We have found 51 new QSOs
(4 BAL QSOs), 13 Seyfert Galaxies (5 NLSy1's), 23 emission line galaxies, 
3 BL Lac objects and 57 stars.

\end{abstract}

\section{Introduction}
 
Optically selected quasar samples are the most complete ones. However, every
survey technique has a redshift and a luminosity-dependent selection biases 
(Wampler and Ponz 1985). Only the combination of the different search 
techniques in different spectral ranges, will yield a complete sample of 
quasars, with an adequate representation of the properties inherent to
them (Hartwick \& Shade 1990; Chavushyan 1995). We have started a 
multiwavelength search for QSOs in the well optically investigated SBS sky area. 
(Stepanian et al. 1999, and references therein). 

 The FIRST radio survey (Becker et al. 1995) provides a new resource for
constructing a large quasar sample. With positions accurate to
1\arcsec ~and a point source sensitivity limit of 1 mJy, it goes 50 times 
fainter in flux than any previous radio survey. The FBQS (Gregg et al. 1996; 
White et al. 2000) is a QSO survey based on FIRST. Unfortunately, for the FBQS 
area there is a lack of complete optically selected samples of QSOs. Hence,
it is very difficult to estimate the completeness of this latter survey, as well as, 
the number of missed quasars in the optical FBQS sample.

\section{Sample}

In order to detect the missing QSOs in the SBS, and to create
a more complete QSO sample, we have cross-identified the deep radio (FIRST) 
and optical (APM) databases. The overlap of the surveys is about 700 sq. degree 
in the sky. 
We have selected objects classified as stellar-like on APM with B-magnitudes 
between 14\fm5 and 18\fm5 coincident with the positions of the FIRST sources 
within 3\arcsec ~radius. The sources, in this 
region, are $\sim 400$ objects, of which 90 were 
previously known AGNs (discovered mainly by the SBS) (Fig. 1).

\section{Observations and preliminary results}

The observations were carried out with the 2.1m  telescope of the GHO and 
the LFOSC spectrograph. A set-up covering the spectral range of 
4200-9000 \AA,  with a dispersion of 8 \AA/pixel was
adopted (Zickgraf et al. 1997). During 1999-2000 period, we have carried out 
spectroscopic observations for ~150 FAQS objects (see examples in Fig.2). 
So far, in this studied subsample, we have found 51 new QSOs (4 BAL), 13 Seyfert
Galaxies (5 NLSy1's), 23 emission line galaxies, 3 BL Lac objects and 57 high galactic 
latitude stars.

\noindent This work was supported by CONACyT research grants 28499-E, J32178-E and, G28586-E.
   
\begin{figure}
\centering
\includegraphics[width=13cm,clip=]{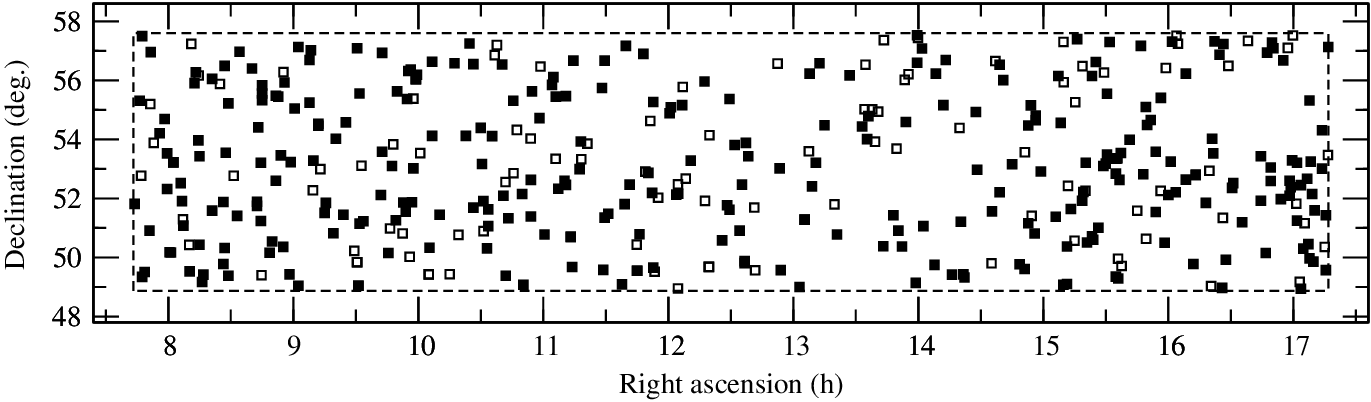}
\caption{The distribution of the FAQS objects on the sky. Open squares
representing new radiosources and filled squares representing the previously
known AGNs.}   \end{figure}

\begin{figure}
\centering
\includegraphics[width=13cm,clip=]{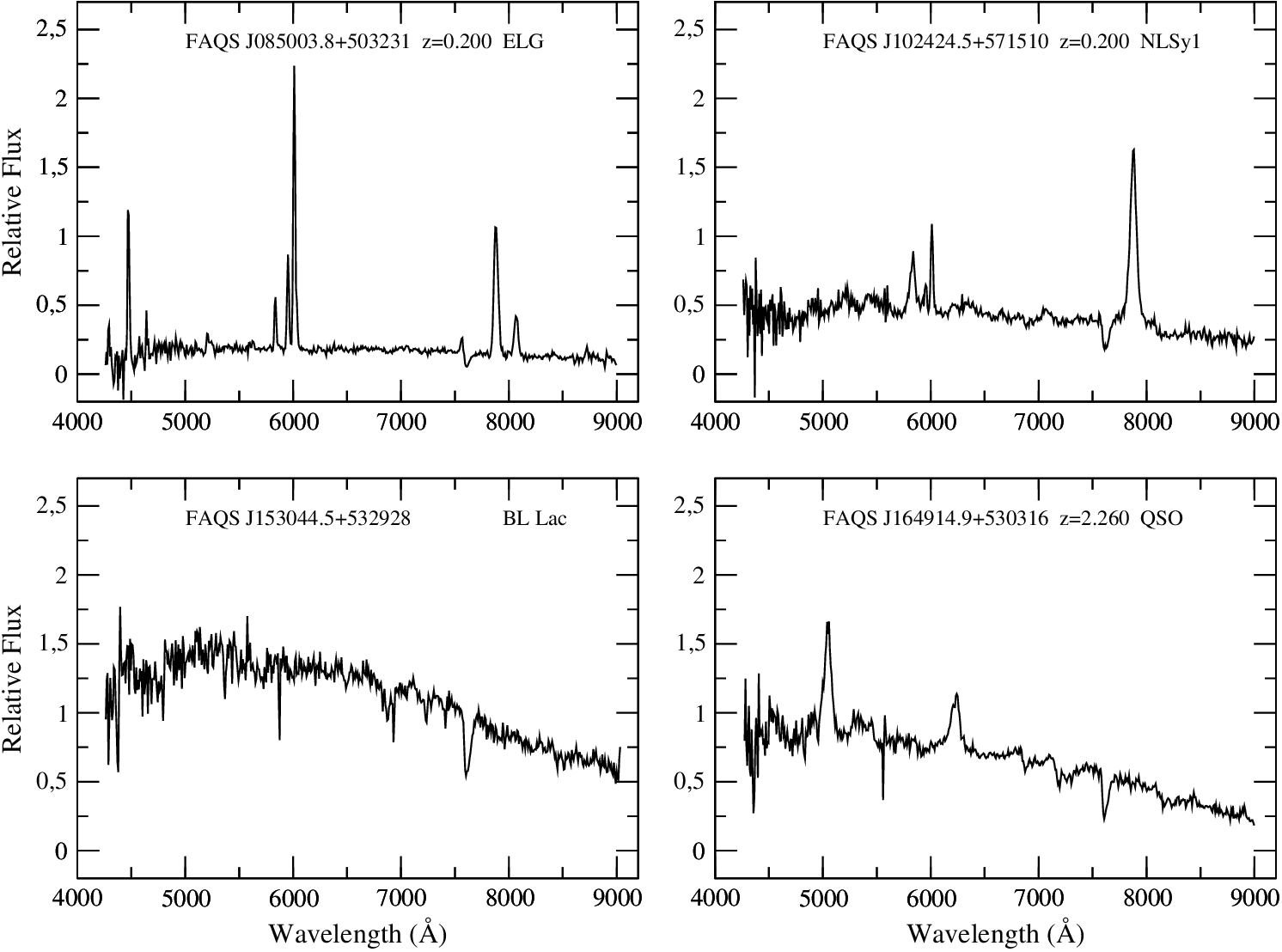}
\caption{Typical examples of LFOSC spectra of different type of objects
contained in the FAQS sample. }  
\end{figure}

\begin{figure}
\centering \includegraphics[width=13cm,clip=]{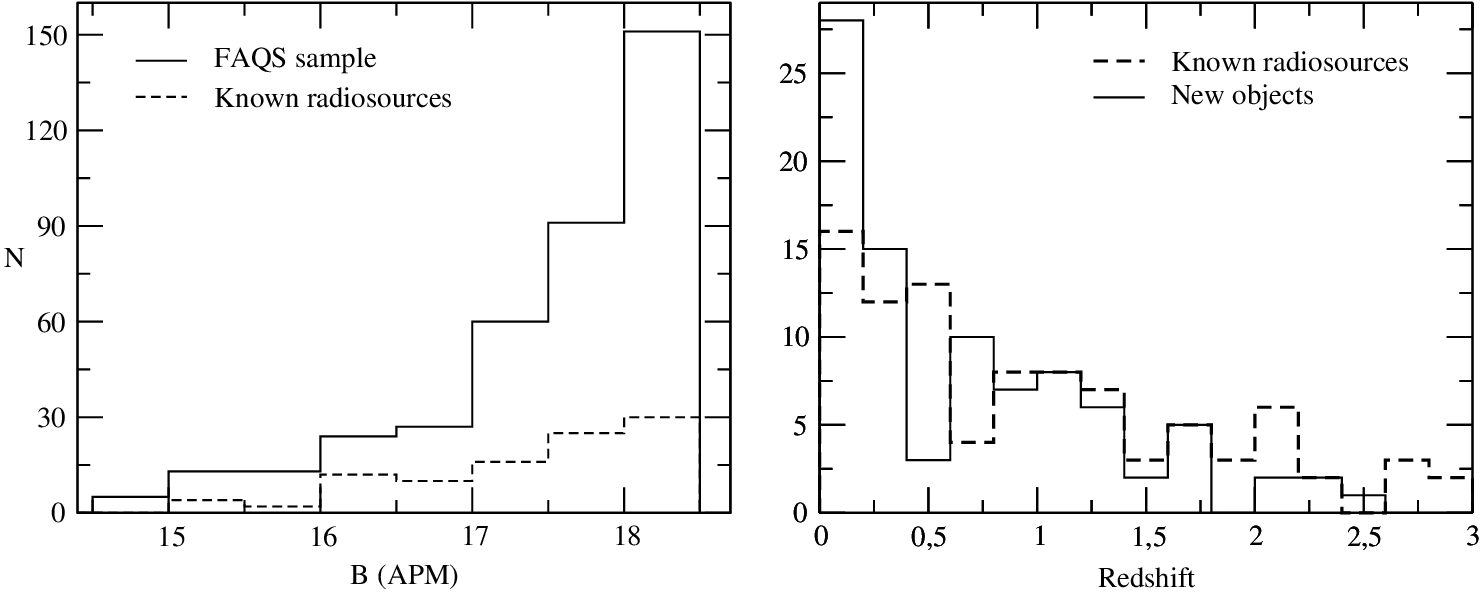}
\par \caption{Histograms of the magnitude and redshift
distributions for the FAQS sample.}
\end{figure}

\end{document}